\documentclass[
 reprint,
 showpacs, preprintnumbers,
 amsmath, amssymb,
 aps,
 pra,
 floatfix,
 superscriptaddress%,
]{revtex4-2}

\usepackage{graphicx}
\usepackage{tikz}
\usepackage{xcolor}
\usetikzlibrary{arrows}

\makeatletter
\newcommand{\colorcaption}[2][]{%
  \begingroup%
  \renewcommand{\@caption@fignum@sep}{ (Color online). }%
  \caption[#1]{#2}%
  \endgroup%
}
\makeatother

\makeatletter
\def\maketag@@@#1{\hbox{\m@th\normalfont\normalsize#1}}
\makeatother

\usepackage{dcolumn}
\usepackage{bm}
\usepackage{bbm}
\usepackage{upgreek}
\usepackage{hyperref}
\usepackage{blindtext}
\usepackage{hyphsubst}

\usepackage{changes}

\usepackage{dsfont}

\usepackage{newtxtext,newtxmath}

\definecolor{darkgreen}{RGB}{0 100 0}

\usepackage{pdfpages}
\makeatletter
\AtBeginDocument{\let\LS@rot\@undefined}
\makeatother

\begin{document}

	\preprint{}

    %%-------------------------------------------------------------------------------------------

	\title{Unconventional Josephson supercurrent diode effect induced by chiral spin-orbit coupling}
	
	\author{Andreas Costa}%
	\email[Corresponding author: ]{andreas.costa@physik.uni-regensburg.de}
 	\affiliation{Institute for Theoretical Physics, University of Regensburg, 93040 Regensburg, Germany}

    \author{Osamu Kanehira}%
 	\affiliation{Department of Applied Physics, Tohoku University, Sendai 980-8579, Japan}

 	\author{Hiroaki Matsueda}%
    \affiliation{Department of Applied Physics, Tohoku University, Sendai 980-8579, Japan}
 	\affiliation{Center for Science and Innovation in Spintronics, Sendai 980-8577, Japan}
 	 	
 	\author{Jaroslav Fabian}%
 	\affiliation{Institute for Theoretical Physics, University of Regensburg, 93040 Regensburg, Germany}

 	\date{\today}

    %%-------------------------------------------------------------------------------------------
    
    \begin{abstract}
        Chiral materials lacking mirror symmetry can exhibit unconventional spin-orbit fields, including fully momentum-aligned radial Rashba fields as seen in twisted van der Waals homobilayers. We theoretically study Cooper-pair transfer in superconductor/ferromagnet/superconductor Josephson junctions with crossed (tangential and radial) interfacial Rashba fields. We find that their interplay leads to what we call the unconventional supercurrent diode effect~(SDE), where supercurrent rectification occurs even with collinear (with respect to the current) barrier magnetization, not possible for conventional spin-orbit fields. This SDE, distinct from conventional Rashba-induced effects on Cooper-pair momenta, arises from the spin precession in the magnetic barrier. We propose it as a sensitive probe of chiral spin textures.
    \end{abstract}
    
    %%-------------------------------------------------------------------------------------------
    
    \maketitle
    
    %%-------------------------------------------------------------------------------------------

    {\color{blue} \emph{Introduction. }}%%
    The interplay of spin-orbit coupling~(SOC) and magnetism is essential for spintronics applications~\cite{Fabian2004,Fabian2007}, enabling transport magnetoanisotropies~\cite{Moser2007,MatosAbiague2009}, tunneling Hall effects~\cite{MatosAbiague2015,Rylkov2017}, or spin-orbit torque in magnetic tunnel junctions~\cite{Gao2020}. 
    Particularly sensitive probes are superconducting junctions~\cite{Eschrig2011,Linder2015}, in which superconducting coherence can significantly amplify this interplay~\cite{Hoegl2015,Hoegl2015,*Hoegl2015a,Costa2017,Costa2019, Martinez2020, Vezin2020} and lead, e.g., to triplet pairing~\cite{Bergeret2001,Volkov2003,Keizer2006,Halterman2007,Eschrig2008,Eschrig2011,Sun2015,Costa2021}.
    
    In Josephson junctions formed by Al/InAs heterostructures, the interfacial Rashba field induces supercurrent rectification in the presence of an in-plane magnetic field perpendicular to the current direction~\cite{Baumgartner2022a,Baumgartner2022,Jeon2022,Pal2022,Turini2022,FrolovQW2022,Costa2023,Costa2023a,Banerjee2023,Kochan2023,Banerjee2024a,Kokkeler2024,Reinhardt2024,Scharf2024,Kang2024}. 
    This supercurrent diode effect~(SDE) has been observed in a variety of systems including superconducting superlattices~\cite{Ando2020}, twisted bi- and trilayer graphene~\cite{DiezMerida2023,lin2022zerofield,scammell2022theory}, 
     der Waals heterostructures~\cite{Wu2022,Bauriedl2022,Yun2023,Kim2024},
    or topological materials~\cite{BoLu2022,Pal2022,Fu2022TopoDiode,Fracassi2024}. 
    Detecting the SDE in twisted multilayer high-temperature superconductors~\cite{Can2021,Zhao2023,Ghosh2024,Volkov2024,Yerin2024} indicates unconventional (e.g., $ d $-wave-like) superconducting pairing.

    The common argument for the SDE is the formation of Cooper pairs with a finite center-of-mass momentum due to the distorted Fermi surface in the presence of SOC, such as Rashba~\cite{Bychkov1984,Bychkov1984b,*Bychkov1984c} and Dresselhaus~\cite{Dresselhaus1955}, and an in-plane magnetic field~\cite{Daido2022,Yuan2022,He2022,Ilic2022,Davydova2022,Banerjee2023}. 
    This results in marked phase asymmetries of the Andreev states~\cite{Andreev1966,*Andreev1966alt} and anomalous $ \varphi_0 $-phase shifts~\cite{Bezuglyi2002,Krive2004,Buzdin2008,Reynoso2008,Grein2009,Zazunov2009,Liu2010a,Liu2010,Liu2011,Reynoso2012,YokoyamaJPSJ2013,Brunetti2013,Shen2014,Yokoyama2014Anomalous,Konschelle2015,Szombati2016,Assouline2019,Mayer2020b,Strambini2020} in the current--phase relations~(CPRs) of the individual transverse channels. 
    The interference of multiple channels in wide junctions---each with slightly different $ \varphi_0 $---leads to a strongly distorted total CPR and a sizable Josephson SDE~\cite{Baumgartner2022a,Baumgartner2022,Costa2023,Costa2023a,Lotfizadeh2023,Reinhardt2024,Patil2024}. 
    Recent theoretical works~\cite{Cheng2023,Debnath2024} investigating supercurrents through a chiral quantum dot argue for an SDE even without SOC and finite Cooper-pair momentum. 

    Rashba and Dresselhaus spin-orbit fields in magnetic junctions are well studied, particularly in III--V semiconductors~\cite{Rozhansky2008,Rozhansky2016} like InAs being used in Al/InAs/Al Josephson junctions~\cite{Baumgartner2022a,Baumgartner2022}.     
    However, the recently discovered chiral spin textures in topological materials~\cite{Bradlyn2016,Schroeter2019,Sakano2020,MeraAcosta2021,Calavalle2022,Gosalbez2023,Krieger2024} and predictions of purely radial Rashba SOC in twisted heterostructures~\cite{Frank2024,Menichetti2023} offer ways for controlling magnetotransport, as shown by large-scale simulations of quantum focusing~\cite{Kang2024}.

    In this Letter, we investigate Josephson junctions with a magnetic barrier and interfacial regions that feature crossed conventional~(CR) and radial Rashba~(RR) fields in the plane transverse to the tunneling. 
    While it is expected that a magnetization in the plane of the spin-orbit fields leads to the SDE, for the crossed fields we find a marked nonreciprocal critical current for the magnetization of the barrier collinear with the transport direction and perpendicular to the spin-orbit fields. 
    We term this effect \emph{unconventional SDE}~(USDE). 
    The mechanism for the USDE is polarity- and field-orientation-dependent precession of the electron spins, conditioned by the interfacial spin-orbit fields, by the magnetization of the tunneling barrier finally resulting in different transmission probabilities for left- and right-propagating electrons. 
    The USDE is different from the commonly considered finite-momentum Cooper-pair generation in the conventional SDE, which requires the magnetization (or an external magnetic field) to lie in the plane of the Rashba field (if the magnetization is perpendicular, as in our case, the Cooper pairs have nominally zero momentum in each electrode). Our numerical model calculations illustrate that the USDE is sizable already at weak radial Rashba coupling, reproducing all symmetries (tunabilities) with respect to the out-of-plane magnetization orientation and Rashba angle that are predicted by our spin-precession picture.

\begin{figure}
        \centering
        \includegraphics[width=0.475\textwidth]{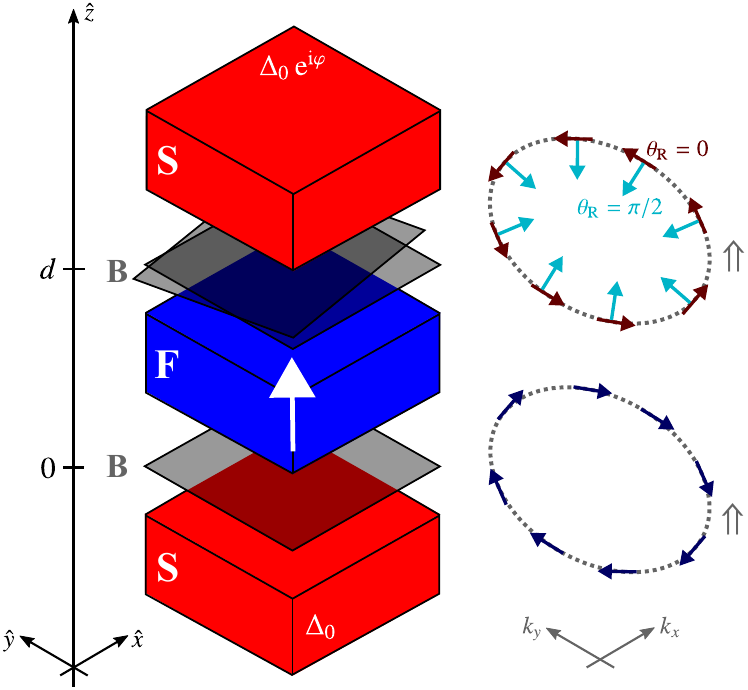}
        \caption{Sketch of the vertical S/F/S Josephson junction using $ C_{2v} $ principal crystallographic orientations $ \hat{x} \parallel [110] $, $ \hat{y} \parallel [\overline{1}10] $, and $ \hat{z} \parallel [001] $; $ \Delta_0 $ ($ \varphi $) indicates the superconducting gap (phase difference) and the magnetization direction (white arrow) of the F with length $ d $ aligned along $ +\hat{z} $~(out of plane). The two-dimensional monolayer barrier~(B) at $ z=0 $ induces CR SOC~(dark-blue arrows, shown along the spin-up Fermi surface $ \Uparrow $), while the twisted bilayer B at $ z=d $ results in crossed CR~(dark-red) and RR~(turquoise) SOCs depending on the Rashba angle~$ \theta_\mathrm{R} $. }
        \label{fig:1}
    \end{figure}

    {\color{blue} \emph{Theoretical model. }}%%
    We consider an epitaxially grown, ballistic, vertical superconductor/ferromagnet/superconductor (S/F/S) Josephson junction with two semi-infinite $ s $-wave superconducting electrodes~(S; spanning $ z<0 $ and $ z>d $) weakly coupled by a ferromagnet~(F; spanning $ 0 < z < d $) and that hosts ultrathin tunneling barriers at both F/S interfaces, as illustrated in Fig.~\ref{fig:1}. 
    The interface at $ z=0 $ could, e.g., consist of a van der Waals monolayer to induce---apart from scalar potential scattering---CR SOC, while a twisted heterostructure or another chiral material induces an unconventional RR component, quantified by the twist-angle-dependent~\cite{Frank2024} Rashba angle $ \theta_\mathrm{R} $, at the $ z = d $ interface. 
    A possible experimental realization could be based on $ \mathrm{NbSe}_2 $ as superconducting electrodes~\cite{Bauriedl2022} and a twisted heterostructure of the quasi-2D van der Waals magnet $ \mathrm{Fe}_{3-x} \mathrm{Ge} \mathrm{Te}_2 $~\cite{Iturriaga2023} serving as magnetic link and inducing crossed Rashba fields due to its lacking mirror symmetry.

    Solving the stationary Bogoljubov--de Gennes~(BdG)~\cite{DeGennes1989} equation $ \hat{\mathcal{H}}_\mathrm{BdG} \Psi(\mathbf{r}) = E \Psi(\mathbf{r}) $, with the corresponding BdG Hamiltonian 
    \begin{equation}
        \hat{\mathcal{H}}_\mathrm{BdG} = \left[ \begin{matrix} \hat{\mathcal{H}}_\mathrm{e} & \hat{\Delta}_\mathrm{S}(z) \\ \hat{\Delta}_\mathrm{S}^\dagger(z) & \hat{\mathcal{H}}_\mathrm{h} \end{matrix} \right] , 
        \label{eq:BdG}  
    \end{equation}
    yields the general scattering states~$ \Psi(\mathbf{r}) $ for incoming spin-polarized electronlike and holelike quasiparticles of excitation energy~$ E $, from which we obtain the Andreev-reflection amplitudes. 
    These coefficients provide the input to numerically compute the CPRs $ I=I(\varphi) $ through the Green's function-based Furusaki--Tsukada approach~\cite{Furusaki1991}, as outlined in the Supplemental Material~(SM)~\footnote{See the Supplemental Material---including Refs.~\cite{DeGennes1989,Furusaki1991,Costa2017,Furusaki1994,Groth2014,Zuo2017,Daido2022,Yuan2022,He2022,Ilic2022,Davydova2022,Banerjee2023}---at [link] for more details.}. 
    Assuming equal Fermi levels $ \mu $ and effective masses $ m $ throughout the junction, the effective Hamiltonian for unpaired electrons reads as $ \hat{\mathcal{H}}_\mathrm{e} = [ -\hbar^2/(2m) \boldsymbol{\nabla}^2 - \mu ] \hat{\sigma}_0 - (\Delta_\mathrm{XC}/2) \Theta(z) \Theta (d-z) (\hat{\mathbf{m}} \cdot \hat{\boldsymbol{\sigma}}) + \hat{\mathcal{H}}_\mathrm{B} $, its holelike counterpart reads as $ \hat{\mathcal{H}}_\mathrm{h} = -\hat{\sigma}_y \hat{\mathcal{H}}_\mathrm{e}^* \hat{\sigma}_y $, and the S pairing potential is approximated~\cite{Beenakker1997} by $ \hat{\Delta}_\mathrm{S}(z) = \Delta_0 \tanh(1.74 \sqrt{T_\mathrm{c}/T - 1}) [ \Theta(-z) + \mathrm{e}^{\mathrm{i} \varphi} \Theta(z-d) ] \hat{\sigma}_0 $ with the zero-temperature gap $ \Delta_0 = 2.5 \, \mathrm{meV} \approx 10^{-3} \mu $, the ratio between temperature~$ T $ and critical temperature~$ T_\mathrm{c} $ set to~$ T/T_\mathrm{c} = 0.1 $, and the phase difference~$ \varphi $. 
    The unit vector $ \hat{\mathbf{m}} = [ \cos(\Theta) \cos(\Phi) , \cos(\Theta) \sin(\Phi) , \sin(\Theta) ]^\top $ describes the magnetization direction inside the F with exchange splitting $ \Delta_\mathrm{XC} $ in terms of the out-of-plane angle~$ \Theta $~(with respect to the interfaces) and in-plane azimuthal angle $ \Phi $~($ \Phi = 0.5\pi $ in all numerical calculations), while $ \hat{\sigma}_0 $~($ \hat{\sigma}_i $) indicates the $ 2 \times 2 $ identity ($ i $th Pauli spin) matrix and $ \hat{\boldsymbol{\sigma}} = [\hat{\sigma}_x , \hat{\sigma}_y , \hat{\sigma}_z ]^\top $. 
    The ultrathin barriers of height~(width) $ V_\mathrm{B} $~($ d_\mathrm{B} $) are described in the deltalike form $ \hat{\mathcal{H}}_\mathrm{B} = V_\mathrm{B} d_\mathrm{B} \hat{\sigma}_0 [\delta(z) + \delta(z-d)] + \hat{\boldsymbol{\Omega}} (\mathbf{k_\parallel}) \cdot \hat{\boldsymbol{\sigma}} $, with the Rashba spin-orbit field $ \hat{\boldsymbol{\Omega}}(\mathbf{k_\parallel}) = \hat{\boldsymbol{\Omega}}(k_x, k_y ) = \alpha [k_y , -k_x , 0] \delta(z) + \alpha [- \sin(\theta_\mathrm{R}) k_x - \cos(\theta_\mathrm{R}) k_y , \cos(\theta_\mathrm{R}) k_x - \sin(\theta_\mathrm{R}) k_y , 0 ] \delta(z-d) $ accounting for CR~(crossed CR and RR) SOC of strength $ \alpha $ at the $ z=0 $~($ z = d $) interface.

    {\color{blue} \emph{Numerical results. }}%%
    To analyze our numerical results, we introduce the dimensionless parameters $ P = (\Delta_\mathrm{XC}/2)/\mu $~[$ k_\mathrm{F} d $] for the spin polarization~[effective length] of the F, while $ Z = 2m V_\mathrm{B} d_\mathrm{B} / (\hbar^2 k_\mathrm{F}) $ and $ \lambda_\mathrm{R} = 2m \alpha / \hbar^2 $ quantify the interfacial barrier and Rashba strengths. 
    We consider a weak F described by $ P = 0.4 $ and $ k_\mathrm{F} d = 12 $~(corresponding to a length of a few nanometers depending on the Fermi level), as well as high interfacial transparencies of~$ \tau = [1+(Z/2)^2]^{-1} = 80 \, \% $~(equivalently~$ Z=1 $)~\cite{Blonder1982} and realistic Rashba SOC~$ \lambda_\mathrm{R} = 1 $~\cite{Hoegl2015,*Hoegl2015a,Martinez2020}.

    \begin{figure}
        \centering
        \includegraphics[width=0.475\textwidth]{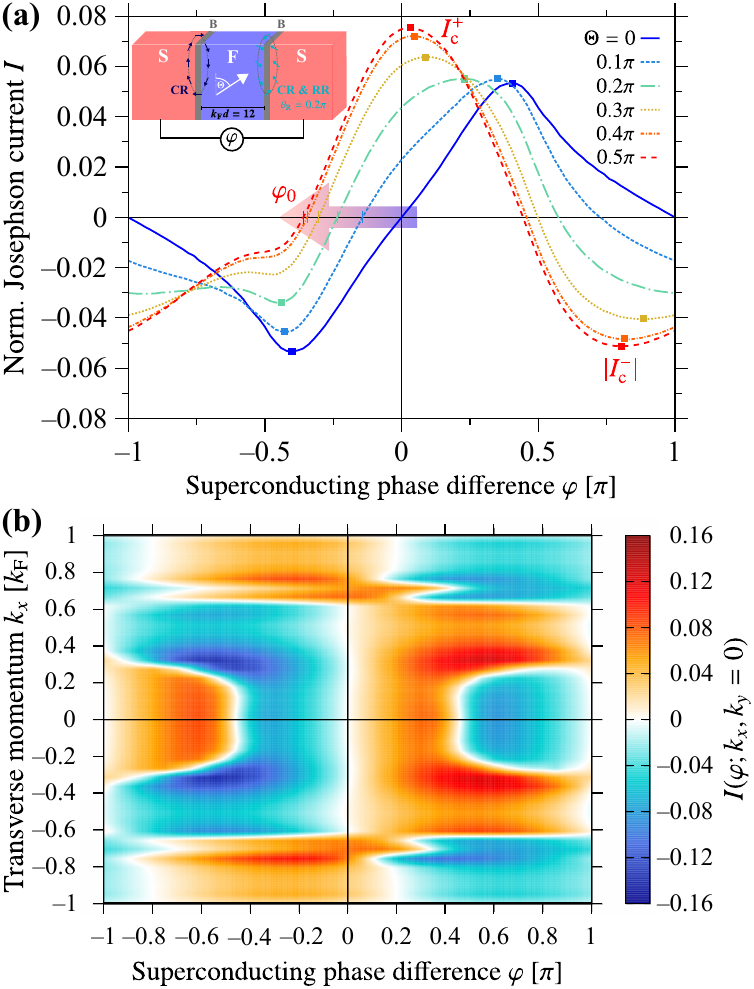}
        \caption{(a)~CPRs $ I(\varphi) $ for the effective F length $ k_\mathrm{F} d = 12 $, Rashba SOC $ \lambda_\mathrm{R} = 2m\alpha / \hbar^2 = 1 $, Rashba angle $ \theta_\mathrm{R} = 0.2 \pi $, and indicated out-of-plane magnetization angles~$ \Theta $; the polarity-dependent critical currents $ I_\mathrm{c}^+ $ and $ |I_\mathrm{c}^-| $ are indicated by squares, and the $ \varphi_0 $ shifts by ticks along the $ \varphi $ axis. (b)~Resolved CPRs $ I(\varphi ; k_x,k_y=0) $ for transverse channels with momenta $ k_x \in [-k_\mathrm{F} ; k_\mathrm{F}] $~(while $ k_y=0 $) and out-of-plane magnetization $ \Theta = 0.5\pi $. The total CPR in (a)---dashed red curve---follows from integrating out $ k_x $ and $ k_y $; white regions indicate the $ \varphi_0 $-phase shifts~(CPR zero crossings). The current is always given in multiples of $ \pi \Delta_0 G_\mathrm{S} / e $; $ G_\mathrm{S} = A e^2 k_\mathrm{F}^2 / (2\pi h) $ is Sharvin's conductance of a point contact with cross section $ A $. }
        \label{fig:2}
    \end{figure}

\begin{figure}
        \centering
        \includegraphics[width=0.475\textwidth]{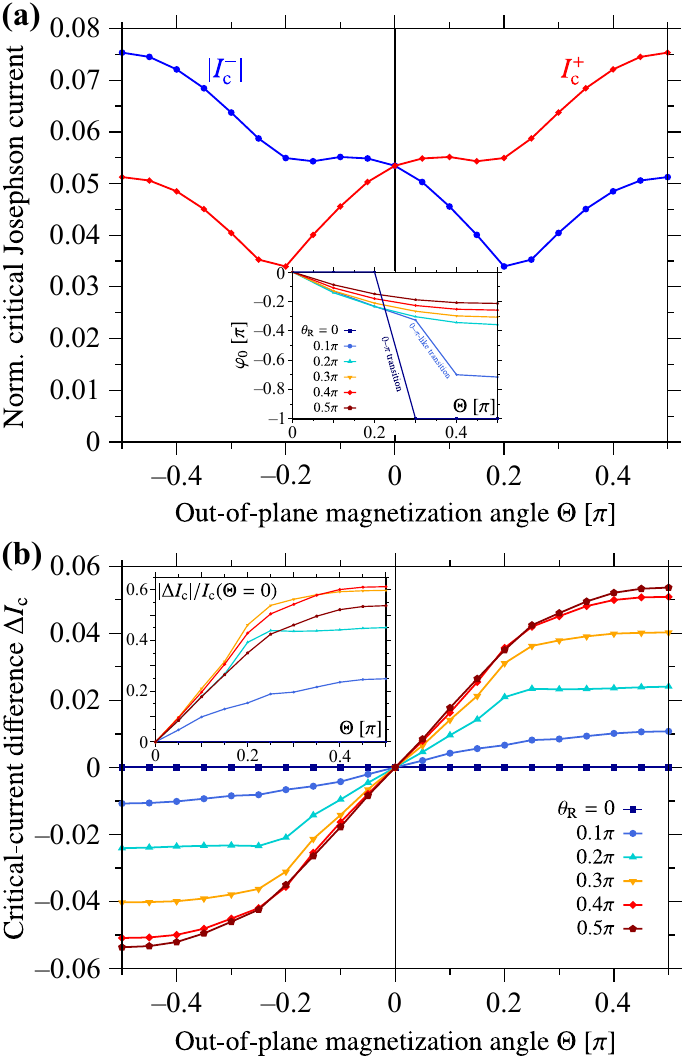}
        \caption{(a)~Polarity-dependent critical-current amplitudes~$ I_\mathrm{c}^+ $ and $ |I_\mathrm{c}^-| $ as functions of the out-of-plane magnetization angle $ \Theta $ and for the same parameters as in~Fig.~\ref{fig:2}(a). Inset: $ \varphi_0 $ shifts \emph{vs.} $ \Theta $ for indicated Rashba angles~$ \theta_\mathrm{R} $. (b)~Dependence of the critical-current difference~$ \Delta I_\mathrm{c} = I_\mathrm{c}^+ - |I_\mathrm{c}^-| $ on $ \Theta $ for indicated $ \theta_\mathrm{R} $; currents are normalized as in~Fig.~\ref{fig:2}. Inset: $ |\Delta I_\mathrm{c}| $ normalized to the polarity-independent critical current~$ I_\mathrm{c}(\Theta = 0) $ for in-plane magnetization. }
        \label{fig:3}
    \end{figure}

    Figure~\ref{fig:2}(a) illustrates a generic case: the CPRs~$ I(\varphi) $ in the presence of CR at the $ z = 0 $ and \emph{crossed} Rashba fields with Rashba angle $ \theta_\mathrm{R} = 0.2\pi $ at the $ z = d $ interface, with variable F magnetization from the in-plane to the positive out-of-plane orientation. 
    If the magnetization is fully aligned in the interfacial plane~($ \Theta = 0 $), the CPR is strictly point symmetric with respect to zero phase difference and the critical-current amplitudes are polarity independent, i.e., $ I_\mathrm{c}^+ = |I_\mathrm{c}^-| \equiv I_\mathrm{c}(\Theta=0) $. 
    With increasing $ \Theta $, a finite out-of-plane component~$ m_z \neq 0 $ is imprinted on the magnetization and the electron spins precess in the F as explained in the next section. 
    The precession with respect to the radial Rashba field leads to polarity-dependent transmission probabilities, manifesting themselves in polarity-dependent critical-current amplitudes $ I_\mathrm{c}^+ \neq |I_\mathrm{c}^-| $ as a direct signature of the USDE. 
    Furthermore, the CPRs acquire intrinsic $ \varphi_0 $ shifts---such that $ I(\varphi) \propto \sin(\varphi - \varphi_0) $ in the simplest case---effectively shifting their zero crossings and inflection points to finite phase differences~\cite{Costa2023,Costa2023a}; $ |\varphi_0| $ monotonically increases with increasing~$ |\Theta| $ and becomes maximal for fully perpendicular magnetization~($ \Theta = 0.5 \pi $). 
    The amplitudes of both $ \varphi_0 $~[reaching about~$ 0.4 \pi $; see Fig.~\ref{fig:3}(a)] and the USDE~[see Fig.~\ref{fig:3}(b)] are sizable already at rather small Rashba angles~(and weak spin polarization).

    Anticipating our spin-precession picture discussed below reveals a fundamental difference between the USDE and the conventional SDE, in agreement with our numerical results. 
    While the conventional SDE results from a superposition of multiple channels' individual CPRs---all with slightly different $ \varphi_0 $ shifts---that finally distorts the total CPR such that the critical currents have different amplitudes (and while the critical currents of the individual channels are polarity independent)~\cite{Costa2023,Costa2023a}, the \emph{USDE occurs already in single-channel junctions} as the direction-dependent transmission probabilities directly produce nonreciprocal CPRs in all channels with finite transverse momentum~$ |\mathbf{k_\parallel}| \neq 0 $~(at $ |\mathbf{k_\parallel}| = 0 $, SOC vanishes). 
    This explains the larger efficiency of the USDE and the observation that its amplitudes are not directly connected to those of the $ \varphi_0 $ shifts---contrary to the conventional SDE in which sudden jumps of $ \varphi_0 $ close to current-reversing $ 0 $--$ \pi $-like transitions are tightly bound to peaks in the SDE efficiency~\cite{Costa2023,Costa2023a}. 
    For better illustration, the color map in Fig.~\ref{fig:2}(b) shows the channel-resolved~(as a function of $ k_x $; $ k_y = 0 $ for simplicity) CPRs~$ I(\varphi ; k_x , k_y=0) $. 
    Inspecting the color scale, we conclude that $ I_\mathrm{c}^+ $ and $ |I_\mathrm{c}^-| $ are indeed (slightly) different in magnitude for individual $ k_x $ channels~(i.e., the colors indicating maximal-amplitude currents are slightly asymmetric around $ \varphi=0 $ at $ |k_x| \neq 0 $), while the $ \varphi_0 $ shifts are mostly caused by channels with large~$ |k_x| $.

    To quantify the USDE, we extract the amplitudes of the polarity-dependent critical currents $ I_\mathrm{c}^+ $ and $ |I_\mathrm{c}^-| $ from the CPRs to compute the critical-current difference $ \Delta I_\mathrm{c} = I_\mathrm{c}^+ - |I_\mathrm{c}^-| $ as the figure of merit of the SDE. 
    The critical currents are presented as functions of the out-of-plane magnetization angle~$ \Theta $ and for the same Rashba angle as before~($ \theta_\mathrm{R} = 0.2\pi $) in~Fig.~\ref{fig:3}(a). 
    Note that $ I_\mathrm{c}^+ [m_z \propto - \sin(\Theta)] = |I_c^- (-m_z)| $, as anticipated from our spin-precession picture below, holds. 
    While $ I_\mathrm{c}^+ $ monotonically increases with a more dominant out-of-plane $ +\hat{z} $ magnetization~($ 0 < \Theta \leq 0.5\pi $)---which could be a possible signature of a more sizable triplet supercurrent induced by RR SOC---$ |I_\mathrm{c}^-| $ initially decreases, drops into a dip at $ \Theta \approx 0.2 \pi $, and finally increases as well. 
    The reason for the nonmonotonic $ |I_\mathrm{c}^-| $ dependence is the remainder of a current-reversing $ 0 $--$ \pi $(-like) transition, which is most pronounced if both interfaces induce CR SOC~($ \theta_\mathrm{R} = 0 $)---see also Refs.~\cite{Costa2017} and \cite{Note1}---and gets strongly suppressed by a RR component~($ \theta_\mathrm{R} > 0 $) that favors a $ 0 < |\varphi_0| < \pi $ shift instead of a constant $ \pi $ shift; the $ 0 $--$ \pi $(-like) transitions emerge as sudden $ \varphi_0 $ jumps~\footnote{The $ \varphi_0 $ shifts are defined as the superconducting phase difference at which the Josephson energy $ E_\mathrm{J} = \int_0^\varphi \mathrm{d} \phi \, I(\phi) $ is minimized.} in the dark- and light-blue curves~($ \theta_\mathrm{R} = 0 $ and $ \theta_\mathrm{R} = 0.1\pi $) in the inset of Fig.~\ref{fig:3}(a).

    The corresponding dependence of the critical-current difference~$ \Delta I_\mathrm{c} $ on $ \Theta $ is illustrated in~Fig.~\ref{fig:3}(b) tuning the spin-orbit field at the $ z=d $ interface from purely CR~($ \theta_\mathrm{R}=0 $) to purely RR~($ \theta_\mathrm{R}=0.5\pi $) SOC.  
    While $ \Delta I_\mathrm{c} $ scales (nearly perfectly) sinusoidally with $ \Theta $ at large Rashba angles~($ \theta_\mathrm{R} = 0.4 \pi $ and $ 0.5 \pi $), the aforementioned $ 0 $--$ \pi $(-like) transitions may still cause reminiscent deviations at smaller~$ \theta_\mathrm{R} $ values---such as a suddenly emerging steeper increase at small~$ |\Theta| $ followed by a saturation into the maximal $ |\Delta I_\mathrm{c}| $ already at rather small $ |\Theta| $, as seen, e.g., in the turquoise curve for $ \theta_\mathrm{R} = 0.2 \pi $. 
    In agreement with our spin-precession picture below, the maximal $ |\Delta I_\mathrm{c} |$ is reached at fully perpendicular magnetization. 
    Moreover, $ |\Delta I_\mathrm{c}| $ increases monotonically with the Rashba angle~$ \theta_\mathrm{R} $~(the maximal $ |\Delta I_\mathrm{c}| $ at $ \Theta = 0.5\pi $ increases nearly sinusoidally with $ \theta_\mathrm{R} $~\cite{Note1}), indicating that a maximal asymmetry between the SOCs---fully CR at one and fully RR at the other interface---is most effective to maximize the USDE. 
    The sign reversal of $ \Delta I_\mathrm{c} $ when reversing the out-of-plane magnetization direction reflects again that $ I_\mathrm{c}^+ [m_z \propto - \sin(\Theta)] = |I_c^- (-m_z)| $. 
    As pointed out above, and contrary to the conventional SDE, a large USDE does not necessarily coincide with sizable $ |\varphi_0| $-CPR shifts. 
    Figure~\ref{fig:3} confirms this expectation, as the absolute SDE measure $ |\Delta I_\mathrm{c}| $ increases with increasing $ \theta_\mathrm{R} $, while the corresponding $ |\varphi_0| $ simultaneously decreases as a result of the initially present and with increasing $ \theta_\mathrm{R} $ quickly suppressed $ 0 $--$ \pi $(-like) transitions~\footnote{As another cross-check of our simulations, we reversed the orientation of the RR field~(considering, e.g., $ \theta_\mathrm{R} = -0.5 \pi $ instead of $ \theta_\mathrm{R} = 0.5 \pi $) and obtained a sign change of $ \Delta I_\mathrm{c} $ similarly to reversing the out-of-plane magnetization $ m_z $. 
    This observation agrees well with our spin-precession picture, in which flipping the RR-SOC orientation interchanges the precession angles $ \overline{\phi}_\mathrm{RR} $ and $ \phi_\mathrm{RR} $, and acts therefore analogously to reversing $ m_z $. }.

    To give a relative estimate of the USDE, we normalize $ |\Delta I_\mathrm{c}| $ to the polarity-independent critical current for in-plane~(along $ +\hat{y} $) magnetization $ I_\mathrm{c}(\Theta=0) $ in~Fig.~\ref{fig:3}(b)~\footnote{Note that we need to define a different relative measure for the USDE efficiency than for the conventional SDE in lateral Josephson junctions~\cite{Costa2023}, in which the effect is induced by an applied in-plane magnetic field and $ |\Delta I_\mathrm{c}| $ is typically compared to the critical current after turning off the field. Regarding the USDE studied here, the exchange splitting in the F is always present in the experiment and the polarity-independent critical currents at spin polarization $ P=0 $ cannot be measured in the same sample. We therefore normalize $ |\Delta I_\mathrm{c}| $ to the polarity-independent critical current~$ I_\mathrm{c}(\Theta=0) $, which is experimentally accessible when rotating the magnetization into the plane. }. 
    Note that $ I_\mathrm{c}(\Theta=0) $ itself depends on $ \theta_\mathrm{R} $, which explains the nonmonotonic dependence of $ |\Delta I_\mathrm{c}| / I_\mathrm{c}(\Theta = 0) $ on $ \theta_\mathrm{R} $. 
    Relative SDE efficiencies beyond $ 20 \, \% $ at small RR SOC and reaching maxima of about $ 60 \, \% $ are sizable, also compared to the conventional SDE~\cite{Costa2023,Costa2023a}.

\begin{figure}
        \centering
        \includegraphics[width=0.475\textwidth]{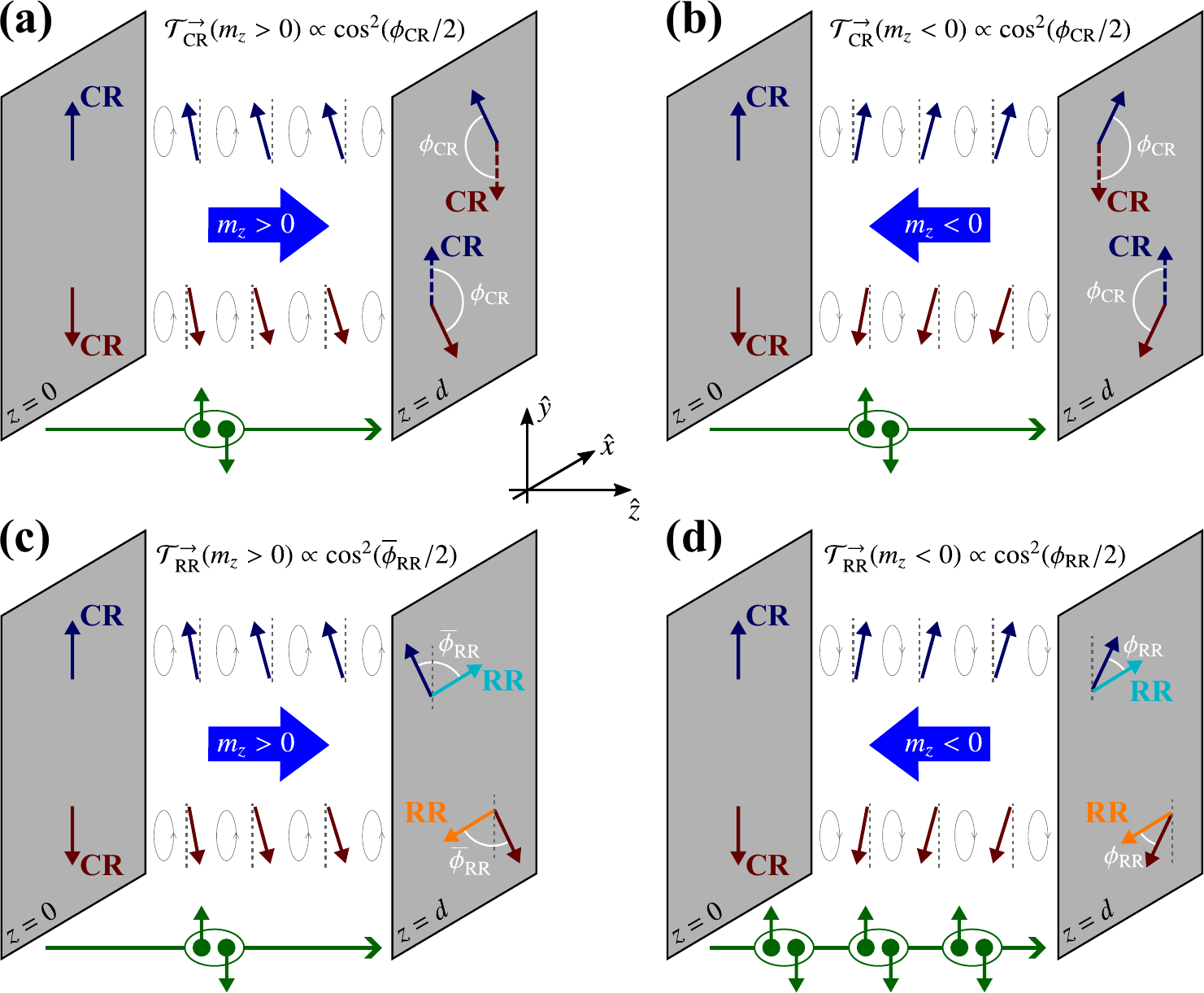}
        \caption{(a)~Spin-resolved electron tunneling~(incident from the left) through the F link~(center) with magnetization $ m_z > 0 $. The spins polarized along $ \pm \hat{y} $ by CR SOC at the $ z=0 $ interface precess counter-clockwise in plane inside the F such that the CR field at the $ z=d $ interface looks effectively rotated by the angle~$ \phi_\mathrm{CR} $; the transmission probability is $ \mathcal{T}_\mathrm{CR}^\rightarrow(m_z>0) $. (b)~Reversing the magnetization~($ m_z<0 $) results in the same tunneling picture with transmission probability $ \mathcal{T}_\mathrm{CR}^\rightarrow(m_z<0) = \mathcal{T}_\mathrm{CR}^\rightarrow(m_z>0) $. The critical-current amplitudes---illustrated by dark-green Cooper pairs---are equal for magnetizations parallel and antiparallel to the current. (c) and (d)~In the presence of RR SOC at the $ z=d $ interface, the spin-precession angles $ \overline{\phi}_\mathrm{RR} \neq \phi_\mathrm{RR} $, as well as the transmission probabilities $ \mathcal{T}_\mathrm{RR}^\rightarrow(m_z > 0) \neq \mathcal{T}_\mathrm{RR}^\rightarrow(m_z < 0) $, depend on the relative orientation between magnetization and current. The critical currents become polarity dependent (Josephson USDE). }
        \label{fig:4}
    \end{figure}

    {\color{blue} \emph{Physical picture. }}%%
    To trace the origin of the USDE in our Josephson junction, we analyze the transmission probabilities of spin-polarized electrons through the junction in the presence of CR SOC at both interfaces~(the SOC field at the $ z = d $ interface is aligned oppositely owing to hybridization), as illustrated in Figs.~\ref{fig:4}(a) and \ref{fig:4}(b). 
    For simplicity, we focus on a single transverse channel with transverse momentum~$ \mathbf{k_\parallel} = [-k_\mathrm{F} , 0]^\top $~($ k_\mathrm{F} = \sqrt{2m\mu} / \hbar $ is the Fermi wave vector), noting that a similar mechanism applies to all other channels.

    The SOC at the $ z=0 $ interface polarizes the spins of the electrons traveling along the $ +\hat{z} $ direction in plane along $ \pm \hat{y} $. 
    When entering the F, the spins are exposed to the magnetization~$ m_z>0 $ pointing along the $ +\hat{z} $ (out-of-plane) direction---i.e., parallel to the electrons' propagation direction---and precess counter-clockwise in the plane parallel to the interfaces. 
    Arriving at the $ z=d $ interface, the precessing spins will effectively see the second spin-orbit field rotated by the angle $ \phi_\mathrm{CR} $ determining their transmission probability $ \mathcal{T}_\mathrm{CR}^\rightarrow \propto \cos^2(\phi_\mathrm{CR}/2) $. 
    As this mechanism applies analogously to spin-up and spin-down electrons, the Cooper-pair-forming electrons sequentially tunnel from the {left ($ z < 0 $) into the right ($ z>d $)} superconductor in the same way, and the corresponding critical Josephson current~$ |I_\mathrm{c}^-(m_z>0)| $ must also scale with~$ \mathcal{T}_\mathrm{CR}^\rightarrow $. 
    For electrons propagating along the opposite $ -\hat{z} $ direction, the tunneling picture is similar~(i.e., $ \mathcal{T}_\mathrm{CR}^\leftarrow = \mathcal{T}_\mathrm{CR}^\rightarrow $), suggesting that the positive critical Josephson current is equal in magnitude, $ I_\mathrm{c}^+(m_z>0) = |I_\mathrm{c}^-(m_z>0)| $.  
    If the magnetization direction is reversed~($ m_z < 0 $), the spins precess also oppositely~(clockwise) but still enclose the same angle~$ \phi_\mathrm{CR} $ with respect to the spin-orbit field at the $z=d $ interface. 
    The critical-current amplitudes are consequently not affected and we expect $ I_\mathrm{c}^+(m_z<0) = |I_\mathrm{c}^-(m_z<0)| = I_\mathrm{c}^+(m_z>0) = |I_\mathrm{c}^-(m_z>0)| $, meaning that the SDE is absent if both spin-orbit fields are conventional; the same arguments rule out the SDE in the case that both spin-orbit fields are radial. 

    Let us now consider \emph{crossed} Rashba fields. Suppose that the chiral spin texture is exhibited by the $ z=d $ interface.  
    As shown in~Figs.~\ref{fig:4}(c) and \ref{fig:4}(d), and in strong contrast to the previous case, the angles between the precessing spins and the radial spin-orbit field are different for magnetizations along~$ +\hat{z} $~(angle $ \overline{\phi}_\mathrm{RR} $) and $ -\hat{z} $~(angle $ \phi_\mathrm{RR} \neq \overline{\phi}_\mathrm{RR} $). 
    Therefore, the transmission probabilities $ \mathcal{T}_\mathrm{RR}^\rightarrow (m_z>0) \propto \cos^2(\overline{\phi}_\mathrm{RR}/2) \neq \mathcal{T}_\mathrm{RR}^\rightarrow (m_z<0) \propto \cos^2(\phi_\mathrm{RR}/2) $ along the $ +\hat{z} $ direction, and the resulting critical currents $ |I_\mathrm{c}^-(m_z>0)| \neq |I_\mathrm{c}^-(m_z<0)| $, depend on the magnetization direction~(parallel or antiparallel) with respect to the current. 
    Moreover, $ \overline{\phi}_\mathrm{RR} $ and $ \phi_\mathrm{RR} $ are interchanged when the electrons travel instead from right to left~(along the $ -\hat{z} $ direction), entailing $ I_\mathrm{c}^+(m_z>0) = |I_\mathrm{c}^-(m_z<0)| \neq I_\mathrm{c}^+(m_z<0) = |I_\mathrm{c}^-(m_z>0)| $, i.e., transport becomes nonreciprocal and the Josephson USDE occurs. 
    The microscopic origin of the USDE---spin-precession-induced polarity- and field-orientation-dependent transmission probabilities---is thus well distinct from the in-plane Cooper-pair momentum being responsible for the conventional SDE.

    {\color{blue} \emph{Alternative realization in lateral junctions. }}%%  
    As an alternative platform for the USDE, we consider a \emph{two-dimensional lateral} S/F/S Josephson junction with in-plane CR in one and RR SOC in the other S, and the F magnetization aligned perpendicular to the plane. 
    The electron Hamiltonian of this junction in tight-binding formulation---entering the BdG Hamiltonian in Eq.~\eqref{eq:BdG}---is given by $ \mathcal{\hat{H}}_\mathrm{e} = \mathcal{\hat{H}}_{\mathrm{S,L}} + \mathcal{\hat{H}}_{\mathrm{F}} + \mathcal{\hat{H}}_{\mathrm{S,R}} $, with 
    {\footnotesize 
    \begin{align}
        \mathcal{\hat{H}}_{\mathrm{S,L}} &= -t \sum_{\langle i,j\rangle,\sigma} \hat{c}_{i,\sigma}^\dagger \hat{c}_{j,\sigma} - \mu\sum_{j,\sigma} \hat{c}_{j,\sigma}^\dagger \hat{c}_{j,\sigma} - \sum_{j}\left[\Delta_0 \hat{c}_{j,\uparrow}^\dagger \hat{c}_{j,\downarrow}^\dagger +\mathrm{h.c.} \right] \nonumber \\
        &\hspace{36pt}+\mathrm i \alpha\sum_{\mu'=x,y}\sum_{\langle i,j\rangle_{\mu'}}\sum_{\alpha',\beta'}\big(\vec n_{\mu'}^{\mathrm{CR}}\cdot\vec\sigma\big)_{\alpha' \beta'} \hat{c}_{i,\alpha'}^\dagger \hat{c}_{j,\beta'}, \\ 
       \mathcal{\hat{H}}_{\mathrm{S,R}} &= -t\sum_{\langle i,j\rangle,\sigma} \hat{c}_{i,\sigma}^\dagger \hat{c}_{j,\sigma} - \mu\sum_{j,\sigma} \hat{c}_{j,\sigma}^\dagger \hat{c}_{j,\sigma} - \sum_{j}\left[\Delta_0 \mathrm{e}^{\mathrm{i} \varphi} \hat{c}_{j,\uparrow}^\dagger \hat{c}_{j,\downarrow}^\dagger +\mathrm{h.c.} \right] \nonumber \\
        &\hspace{36pt}+\mathrm i \alpha \cos(\theta_{\mathrm{R}})\sum_{\mu'=x,y}\sum_{\langle i,j\rangle_{\mu'}}\sum_{\alpha',\beta'}\big(\vec n_{\mu'}^{\mathrm{CR}}\cdot\vec\sigma\big)_{\alpha' \beta'} \hat{c}_{i,\alpha'}^\dagger \hat{c}_{j,\beta'} \nonumber \\ 
        &\hspace{36pt}+\mathrm i \alpha\sin(\theta_{\mathrm{R}})\sum_{\mu'=x,y}\sum_{\langle i,j\rangle_{\mu'}}\sum_{\alpha',\beta'}\big(\vec n_{\mu'}^{\mathrm{RR}}\cdot\vec\sigma\big)_{\alpha' \beta'} \hat{c}_{i,\alpha'}^\dagger \hat{c}_{j,\beta'}, 
        \intertext{\normalsize and}
        \mathcal{\hat{H}}_{\mathrm{F}} &= -t\sum_{\langle i,j\rangle,\sigma} \hat{c}_{i,\sigma}^\dagger \hat{c}_{j,\sigma} - \mu\sum_{j,\sigma} \hat{c}_{j,\sigma}^\dagger \hat{c}_{j,\sigma} - m_z\sum_{j,\sigma} \sigma \hat{c}_{j,\sigma}^\dagger \hat{c}_{j,\sigma} 
    \end{align}}%%
    representing the single-electron Hamiltonians in the left S~($ x < 0 $), right S~($ x > N_\mathrm{F} $), and F~($ 0 < x < N_\mathrm{F} $), respectively; $ \hat{c}_{i,\sigma}^{(\dagger)} $ annihilates~(creates) a spin-$ \sigma $ electron~[$ \sigma = (-)1 $ for spin (down) up] at lattice site~$ i $ and $ \langle i,j \rangle $ indicates nearest-neighbor hoppings. 
    The hopping amplitude, chemical potential, Rashba strength~(angle), $ s $-wave superconducting pairing potential, and amplitude of the exchange field along the $ \hat{z} \parallel [001] $ direction are denoted by $ t $, $ \mu $, $ \alpha $~($ \theta_\mathrm{R} $), $ \Delta_0 $, and $ m_z $, whereas $ \vec{n}_{x}^{\mathrm{CR}}=(0,1,0),~\vec{n}_{y}^{\mathrm{CR}}=(-1,0,0) $ and $ \vec{n}_{x}^{\mathrm{RR}}=(-1,0,0),~\vec{n}_{y}^{\mathrm{RR}}=(0,-1,0) $ are the Rashba vectors for CR and RR SOCs, accordingly. 
    After numerically implementing this tight-binding BdG description in the \textsc{Kwant} Python package~\cite{Groth2014}, we apply a generalized Green's function technique~\cite{Furusaki1994,Zuo2017} to obtain the Josephson CPRs; see the SM~\cite{Note1} for more details. 
    All calculations are performed for a junction with semi-infinite S leads, $ N_\mathrm{F} = 4 $ lattice sites in the F link, and $ N_y = 100 $ sites along the transverse $ \hat{y} $ direction; the other system parameters are $ \mu = -t $, $ \alpha = 0.4t $, $ \Delta_0 = 0.1t $, $ k_\mathrm{B} T = 0.01t $, and $ t = \hbar^2/(2ma^2)= 238 \, \mathrm{meV} $~(or equivalently lattice spacing $ a = 0.4 \, \mathrm{nm} $ if $ m $ is the free-electron mass).

\begin{figure}
        \centering
        \includegraphics[width=0.475\textwidth]{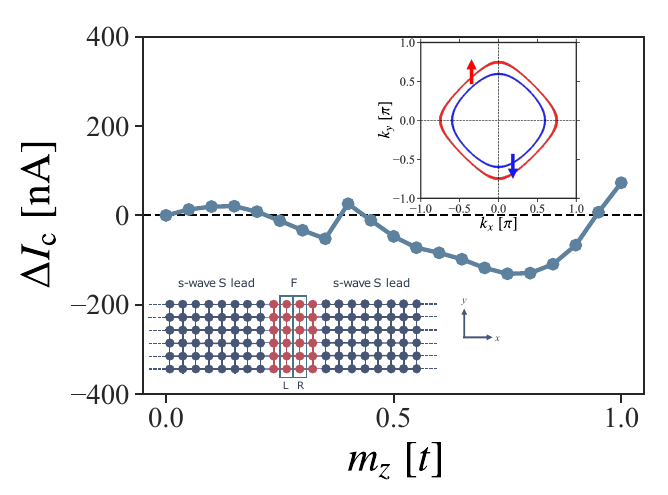}
        \caption{Dependence of the critical-current difference $ \Delta I_\mathrm{c} = I_\mathrm{c}^+ - |I_\mathrm{c}^-| $ of the \emph{lateral} S/F/S junction on the out-of-plane magnetization strength~$ m_z $~(perpendicular to the system plane) at Rashba angle~$ \theta_\mathrm{R} = 0.25\pi $; other parameters are given in the text. The insets show a schematical illustration of the lateral junction in the tight-binding picture~(bottom) and the computed momentum-space spin-up~($ \uparrow $; red) and spin-down~($ \downarrow $; blue) Fermi surfaces in the F link~(top). }
        \label{fig:5}
    \end{figure}

    The calculated critical-current difference $ \Delta I_\mathrm{c} $ for a lateral junction hosting \emph{crossed} Rashba fields~($ \theta_\mathrm{R} = 0.25 \pi $), shown in Fig.~\ref{fig:5}, points to qualitatively similar physics as discussed in the vertical case with the USDE emerging at finite out-of-plane exchange fields $ m_z > 0 $~\cite{Note1}. 
    The maximal USDE~(maximal $ |\Delta I_\mathrm{c}| $) occurs when approaching the half-metallic limit~($ |m_z| \to \mu $)---contrary to vertical S/F/S Josephson junctions in which $ \Delta I_\mathrm{c} $ oscillates as a function of the (effective) length $ k_\mathrm{F} d $ and spin polarization $ P $ of the F link as discussed in the SM~\cite{Note1}---and the pronounced sign reversal of $ \Delta I_\mathrm{c} $ indicates a current-reversing $ 0 $--$ \pi $-like transition~\cite{Costa2023}. 
    The inset of Fig.~\ref{fig:5} illustrates the computed spin-resolved Fermi surfaces inside the F link, which provide no signatures of finite-momentum Cooper pairing~(i.e., no relative shifts or displacements of the Fermi surfaces) and therefore emphasize again that the physical mechanism of the USDE is well distinct from the conventional (finite-momentum-pairing) SDE~\cite{Daido2022,Yuan2022,He2022,Ilic2022,Davydova2022,Banerjee2023}.

    {\color{blue} \emph{Conclusions. }}%%
    In summary, we predicted the Josephson USDE that emerges---contrary to the previously studied conventional SDE---from the interplay of CR and RR SOCs with the out-of-plane magnetization in vertical~(three-dimensional) and lateral~(two-dimensional) S/F/S junctions. 
    In-plane precessions of the current-carrying electrons' spins in the F link trigger different transmission probabilities for left- and right-propagating electrons, as well as for magnetizations parallel and antiparallel to the current, manifesting in nonreciprocal transport and polarity-dependent critical currents. 
    After performing numerical model calculations and analyzing various system parameters, we elaborated more on the qualitative spin-precession picture and deduced its most relevant ramifications on the USDE.  
    We numerically quantified the USDE together with the intrinsic $ \varphi_0 $-CPR shifts, unraveling that the amplitudes of both do not necessarily coincide; while the conventional SDE requires the superposition of many transverse channels with distinct $ \varphi_0 $ to obtain a sizable SDE, the spin-precession mechanism produces the USDE already for a single channel independent of its $ |\varphi_0| $. 
    The efficiency of the USDE is experimentally widely tunable through knobs like the magnetization orientation, Rashba angle, and thickness or spin polarization of the F~\cite{Note1}.

    \begin{acknowledgments}
        A.C. and J.F. gratefully acknowledge funding by the Deutsche Forschungsgemeinschaft (DFG, German Research Foundation) -- Project-IDs 454646522; 314695032. 
        O.K. acknowledges support from GP-Spin at Tohoku University and Japan Science and Technology Agency SPRING Grant No. JPMJSP2114. 
        H.M. acknowledges support from CSIS at Tohoku University. 
    \end{acknowledgments}

    \bibliography{paper}

    \onecolumngrid
    \newpage
    

\section*{Supplementary material}

# SUPPLEMENTAL MATERIAL

# Unconventional Josephson Supercurrent Diode Effect Induced by Chiral Spin-Orbit Coupling

Andreas Costa,[1, *] Osamu Kanehira,[2] Hiroaki Matsueda,[3, 4] and Jaroslav Fabian[1]

[1]*Institute for Theoretical Physics, University of Regensburg, 93040 Regensburg, Germany*
[2]*Department of Applied Physics, Tohoku University, Sendai, Japan*
[3]*Department of Applied Physics, Tohoku University, Sendai 980-8579, Japan*
[4]*Center for Science and Innovation in Spintronics, Sendai 980-8577, Japan*

In this Supplemental Material, we present the technical details of our analytical model for the (three-dimensional) vertical S/F/S Josephson junction considered in the main text, as well as the results of additional model calculations. We furthermore elaborate on a tight-binding model for two-dimensional lateral (planar) S/F/S Josephson junctions and demonstrate that these systems can likewise induce the USDE.

## I. MODEL DETAILS FOR THE VERTICAL S/F/S JOSEPHSON JUNCTION

In the main text, we analytically describe the three-dimensional vertical S/F/S Josephson junction by means of its stationary Bogoljubov–de Gennes Hamiltonian [S1]

$$\hat{\mathcal{H}}_{\text{BdG}} = \begin{bmatrix} \hat{\mathcal{H}}_{\text{e}} & \hat{\Delta}_{\text{S}}(z) \\ \hat{\Delta}_{\text{S}}^{\dagger}(z) & \hat{\mathcal{H}}_{\text{h}} \end{bmatrix}, \tag{S1}$$

where

$$\hat{\mathcal{H}}_{\text{e}} = \left( -\frac{\hbar^2}{2m} \boldsymbol{\nabla}^2 - \mu \right) \hat{\sigma}_0 - \frac{\Delta_{\text{XC}}}{2} \Theta(z) \Theta(d - z) \ (\hat{\mathbf{m}} \cdot \hat{\boldsymbol{\sigma}}) + \hat{\mathcal{H}}_{\text{B}}, \tag{S2}$$

$$\hat{\mathcal{H}}_{\text{h}} = -\hat{\sigma}_y \hat{\mathcal{H}}_{\text{e}}^* \hat{\sigma}_y, \tag{S3}$$

$$\begin{aligned} \hat{\mathcal{H}}_{\text{B}} = V_{\text{B}} \, d_{\text{B}} \, \hat{\sigma}_0 [\delta(z) + \delta(z - d)] \\ + \alpha \, \delta(z) \, [k_y, -k_x, 0] \cdot \hat{\boldsymbol{\sigma}} + \alpha \, \delta(z - d) \, [-\sin(\theta_{\text{R}}) k_x - \cos(\theta_{\text{R}}) k_y, \cos(\theta_{\text{R}}) k_x - \sin(\theta_{\text{R}}) k_y, 0] \cdot \hat{\boldsymbol{\sigma}}, \end{aligned} \tag{S4}$$

and

$$\hat{\Delta}_{\text{S}}(z) = \Delta_0 \, \tanh \left( 1.74 \sqrt{\frac{T_{\text{c}}}{T} - 1} \right) \, [\Theta(-z) + e^{\text{i}\varphi} \Theta(z - d)] \hat{\sigma}_0. \tag{S5}$$

The scattering states $\Psi^{(1)}(\mathbf{r})$ for incident (1) spin-up electronlike quasiparticles of energy $E > 0$ from the left S are obtained as solutions of the Bogoljubov–de Gennes equation $\hat{\mathcal{H}}_{\text{BdG}} \Psi^{(1)}(\mathbf{r}) = E \Psi^{(1)}(\mathbf{r})$, which yields

$$\Psi^{(1)}(\mathbf{r}) = \psi^{(1)}(z) \, e^{\text{i} (\mathbf{k}_{\parallel} \cdot \mathbf{r}_{\parallel})} = \psi^{(1)}(z) \, e^{\text{i} (k_x x + k_y y)} \tag{S6}$$

with

$$\psi^{(1)}(z < 0) = e^{\text{i}q_e z} \begin{bmatrix} u \\ 0 \\ v \\ 0 \end{bmatrix} + \mathcal{A}^{(1)} \, e^{-\text{i}q_e z} \begin{bmatrix} u \\ 0 \\ v \\ 0 \end{bmatrix} + \mathcal{B}^{(1)} \, e^{-\text{i}q_e z} \begin{bmatrix} 0 \\ u \\ 0 \\ v \end{bmatrix} + C^{(1)} \, e^{\text{i}q_h z} \begin{bmatrix} v \\ 0 \\ u \\ 0 \end{bmatrix} + \mathcal{D}^{(1)} \, e^{\text{i}q_h z} \begin{bmatrix} 0 \\ v \\ 0 \\ u \end{bmatrix}, \tag{S7}$$

$$\begin{aligned} \psi^{(1)}(0 < z < d) = \mathcal{E}^{(1)} \, e^{\text{i}k_e^{\uparrow} z} \chi_{\text{e}}^{\uparrow} + \mathcal{F}^{(1)} \, e^{\text{i}k_e^{\downarrow} z} \chi_{\text{e}}^{\downarrow} + \mathcal{G}^{(1)} \, e^{-\text{i}k_h^{\uparrow} z} \chi_{\text{h}}^{\uparrow} + \mathcal{H}^{(1)} \, e^{-\text{i}k_h^{\downarrow} z} \chi_{\text{h}}^{\downarrow} \\ + \mathcal{I}^{(1)} \, e^{-\text{i}k_e^{\uparrow} z} \chi_{\text{e}}^{\uparrow} + \mathcal{J}^{(1)} \, e^{-\text{i}k_e^{\downarrow} z} \chi_{\text{e}}^{\downarrow} + \mathcal{K}^{(1)} \, e^{\text{i}k_h^{\uparrow} z} \chi_{\text{h}}^{\uparrow} + \mathcal{L}^{(1)} \, e^{\text{i}k_h^{\downarrow} z} \chi_{\text{h}}^{\downarrow}, \end{aligned} \tag{S8}$$

---

\* Corresponding author: andreas.costa@physik.uni-regensburg.de

as well as

$$\psi^{(1)}(z > 0) = \mathcal{M}^{(1)} \, \mathrm{e}^{iq_e z} \begin{bmatrix} u \, \mathrm{e}^{i\varphi} \\ 0 \\ v \\ 0 \end{bmatrix} + \mathcal{N}^{(1)} \, \mathrm{e}^{iq_e z} \begin{bmatrix} 0 \\ u \, \mathrm{e}^{i\varphi} \\ 0 \\ v \end{bmatrix} + O^{(1)} \, \mathrm{e}^{-iq_h z} \begin{bmatrix} v \, \mathrm{e}^{i\varphi} \\ 0 \\ u \\ 0 \end{bmatrix} + \mathcal{P}^{(1)} \, \mathrm{e}^{-iq_h z} \begin{bmatrix} 0 \\ v \, \mathrm{e}^{i\varphi} \\ 0 \\ u \end{bmatrix}; \tag{S9}$$

the spinors in the F ($0 < z < d$) are

$$\chi_e^{\uparrow(\downarrow)} = \left[ \chi^{\uparrow(\downarrow)}, 0 \right]^\top \qquad \text{and} \qquad \chi_h^{\uparrow(\downarrow)} = \left[ 0, \chi^{\downarrow(\uparrow)} \right]^\top \tag{S10}$$

with

$$\chi^{\uparrow(\downarrow)} = \frac{1}{\sqrt{2}} \left[ \begin{matrix} (-)\sqrt{1 + (-)\sin(\Theta)} \, \mathrm{e}^{-i\Phi} \\ \sqrt{1 - (+)\sin(\Theta)} \end{matrix} \right]. \tag{S11}$$

Within Andreev approximation ($E, \Delta_0 \ll \mu$), the electron(like) and hole(like) wave vectors are

$$q_e \approx q_h \approx \sqrt{k_F^2 - \mathbf{k}_\parallel^2} \tag{S12}$$

in the S regions ($z < 0$ and $z > d$) and

$$k_e^{\uparrow(\downarrow)} \approx k_h^{\uparrow(\downarrow)} \approx \sqrt{k_F^2[1 + (-)P] - \mathbf{k}_\parallel^2} \tag{S13}$$

in the F ($0 < z < d$); $k_F = \sqrt{2m\mu}/\hbar$ indicates the Fermi wave vector and $P = (\Delta_{XC}/2)/\mu$ the spin polarization of the F. As usual, $u$ and $v$ correspond to the Bardeen–Cooper–Schrieffer coherence factors such that

$$u^2 = \frac{1}{2}\left( 1 + \frac{\sqrt{E^2 - \Delta_0^2}}{E} \right) = 1 - v^2. \tag{S14}$$

Applying the interfacial ($z = 0$ and $z = d$) boundary conditions

$$\psi^{(1)}(z = 0_-) = \psi^{(1)}(z = 0_+), \tag{S15}$$

$$\psi^{(1)}(z = d_-) = \psi^{(1)}(z = d_+), \tag{S16}$$

$$\left\{ \left[ -\frac{\hbar^2}{2m}\frac{\mathrm{d}}{\mathrm{d}z} + V_B d_B \right] \right\} \hat{\eta} \psi^{(1)}(z)\big|_{z=0_+} + \begin{bmatrix} \hat{\mathbf{\Omega}}_\mathbf{L} \cdot \hat{\boldsymbol{\sigma}} & \hat{0}_{2\times2} \\ \hat{0}_{2\times2} & -(\hat{\mathbf{\Omega}}_\mathbf{L} \cdot \hat{\boldsymbol{\sigma}}) \end{bmatrix} \psi^{(1)}(z)\big|_{z=0_+} = -\frac{\hbar^2}{2m}\frac{\mathrm{d}}{\mathrm{d}z} \hat{\eta} \psi^{(1)}(z)\big|_{z=0_-}, \tag{S17}$$

and

$$\left\{ \left[ \frac{\hbar^2}{2m}\frac{\mathrm{d}}{\mathrm{d}z} + V_B d_B \right] \right\} \hat{\eta} \psi^{(1)}(z)\big|_{z=d_-} + \begin{bmatrix} \hat{\mathbf{\Omega}}_\mathbf{R} \cdot \hat{\boldsymbol{\sigma}} & \hat{0}_{2\times2} \\ \hat{0}_{2\times2} & -(\hat{\mathbf{\Omega}}_\mathbf{R} \cdot \hat{\boldsymbol{\sigma}}) \end{bmatrix} \psi^{(1)}(z)\big|_{z=d_-} = \frac{\hbar^2}{2m}\frac{\mathrm{d}}{\mathrm{d}z} \hat{\eta} \psi^\sigma(z)\big|_{z=d_+}, \tag{S18}$$

with

$$\hat{\eta} = \begin{bmatrix} \hat{\sigma}_0 & \hat{0}_{2\times2} \\ \hat{0}_{2\times2} & -\hat{\sigma}_0 \end{bmatrix}, \tag{S19}$$

and the Rashba spin-orbit fields

$$\hat{\mathbf{\Omega}}_\mathbf{L} = \alpha \, [k_y, -k_x, 0] \tag{S20}$$

and

$$\hat{\mathbf{\Omega}}_\mathbf{R} = \alpha \, [-\sin(\theta_R)k_x - \cos(\theta_R)k_y, \cos(\theta_R)k_x - \sin(\theta_R)k_y, 0] \tag{S21}$$

at the left (L) and right (R) junction interfaces, to the scattering states and numerically solving the resulting linear system of equations determines the calligraphically written scattering coefficients $\mathcal{A}^{(1)}, \ldots, \mathcal{P}^{(1)}$. The scattering states $\Psi^{(2)}(\mathbf{r})$, $\Psi^{(3)}(\mathbf{r})$,

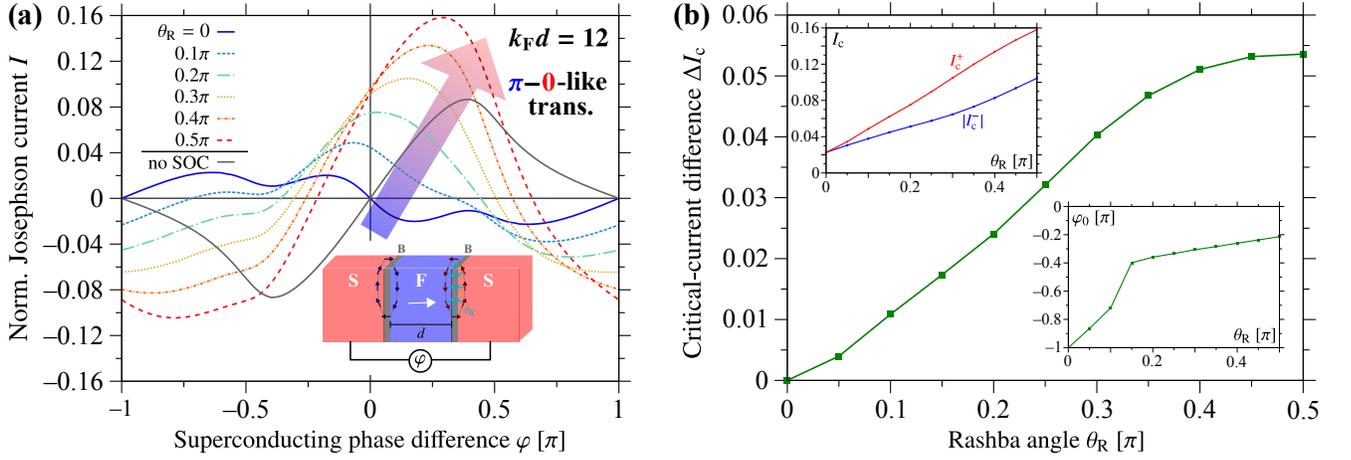

FIG. S1. (a) CPRs $I(\varphi)$—normalized as in the main text—of the vertical S/F/S Josephson junction for the effective F length $k_F d = 12$ and various indicated Rashba angles $\theta_R$. The F magnetization is aligned along the $+\hat{z}$-out-of-plane direction (i.e., $\Theta = 0.5\pi$) and the Rashba strength is $\lambda_R = 2m\alpha/\hbar^2 = 1$. A more dominant RR component (larger $\theta_R$) results in a current-reversing $\pi$–0-like transition. (b) Corresponding critical-current difference $\Delta I_c = I_c^+ - |I_c^-|$ as a function of $\theta_R$. The normalized, polarity-dependent, critical-current amplitudes $I_c^+$ and $|I_c^-|$ are additionally illustrated in the upper inset, while the other inset shows the $\varphi_0$-CPR phase shifts.

and $\Psi^{(4)}(\mathbf{r})$—for incident (2) spin-down electronlike, (3) spin-up holelike, and (4) spin-down holelike quasiparticles—together with the corresponding scattering coefficients—are analogously obtained.

The Josephson CPRs are then computed from the spin-conserving Andreev-reflection coefficients $C^{(1)}$, $D^{(2)}$, $A^{(3)}$, and $B^{(4)}$ according to the Green's function-based Furusaki–Tsukada formula [S2]

$$I = I(\varphi) \approx \frac{\pi\Delta_0 G_S}{e} \frac{k_B T}{4\pi k_F^2} \int d^2\mathbf{k}_\parallel \sum_{\omega_n} \frac{q_e + q_h}{\sqrt{\omega_n^2 + \Delta_0^2}} \left[ \frac{C^{(1)}(i\omega_n) + D^{(2)}(i\omega_n)}{q_e} - \frac{A^{(3)}(i\omega_n) + B^{(4)}(i\omega_n)}{q_h} \right], \qquad (S22)$$

where $G_S = Ae^2 k_F^2/(2\pi h)$ refers to Sharvin's conductance of a three-dimensional point contact with cross section $A$ ($e$ is the positive elementary charge), $k_B T$ is the thermal energy at temperature $T$ (we will consider $T \approx 0.1 T_c$, where $T_c$ is the critical temperature of the S), and $\omega_n = (2n+1)\pi k_B T$ with integer $n$ are the fermionic Matsubara frequencies.

## II. MORE NUMERICAL RESULTS FOR THE VERTICAL S/F/S JOSEPHSON JUNCTION

As we demonstrated in Ref. [S3], interfacial CR SOCs in vertical S/F/S Josephson junctions provide an important knob to induce triplet Cooper pairs and produce sizable supercurrents, even in the maximally spin-polarized half-metallic limit. Nevertheless, the relative phase shift between the two Rashba spin-orbit fields (recall the opposite signs of the Rashba terms at the left and right interfaces owing to hybridization) typically favors the $\pi$-state in the triplet regime, which could be disadvantageous for applications that require switching between 0- and $\pi$-states in a controlled manner.

In the main text, we indeed recovered the SOC-induced $\pi$-state regime if both spin-orbit fields correspond to CR ($\theta_R = 0$) and the magnetization is mostly out-of-plane ($\Theta \geq 0.3\pi$); recall the dark-blue curve in the inset of Fig. 4(a). For predominantly in-plane magnetization ($\Theta \leq 0.2\pi$), the junction is still in the 0-state, but will also transition into the $\pi$-state if the SOC strength $\lambda_R$ is further increased. RR SOC at one of the interfaces (i.e., increasing $\theta_R$), however, introduces an additional relative phase shift between both spin-orbit fields that counteracts the initial $\pi$-shift and switches the junction back into the 0-state.

For illustration, Fig. S1(a) shows the corresponding CPRs for out-of-plane magnetization ($\Theta = 0.5\pi$), effective F thickness $k_F d = 12$, and various Rashba angles $\theta_R$ (similarly to Fig. 3 in the main text, but tuning the Rashba angle $\theta_R$ instead of the magnetization-angle $\Theta$). For CR SOCs at both interfaces ($\theta_R = 0$; blue curve), the junction has undergone a transition into the $\pi$-state when compared to its 0-state in the absence of SOC. Interestingly, already a rather small Rashba angle of $\theta_R = 0.1\pi$ is enough to switch the junction back into the 0-state (as the junction resides, for the considered SOC parameter, very close to the SOC-induced 0–$\pi$ transition), apart from inducing the USDE together with a sizable $\varphi_0$-phase shift that we both analyze in detail in the main text; the $\pi$–0-like back-transition induced by RR SOC is visible as the sudden $\varphi_0$-jump in the inset of Fig. S1(b). Note that the—owing to the USDE polarity-dependent—critical-current amplitudes $I_c^+$ and $|I_c^-|$ both increase monotonically with the Rashba angle $\theta_R$ and finally even exceed the (polarity-independent) critical current in the absence of SOC, possibly indicating the formation of additional triplet Cooper pairs carrying the Josephson current,

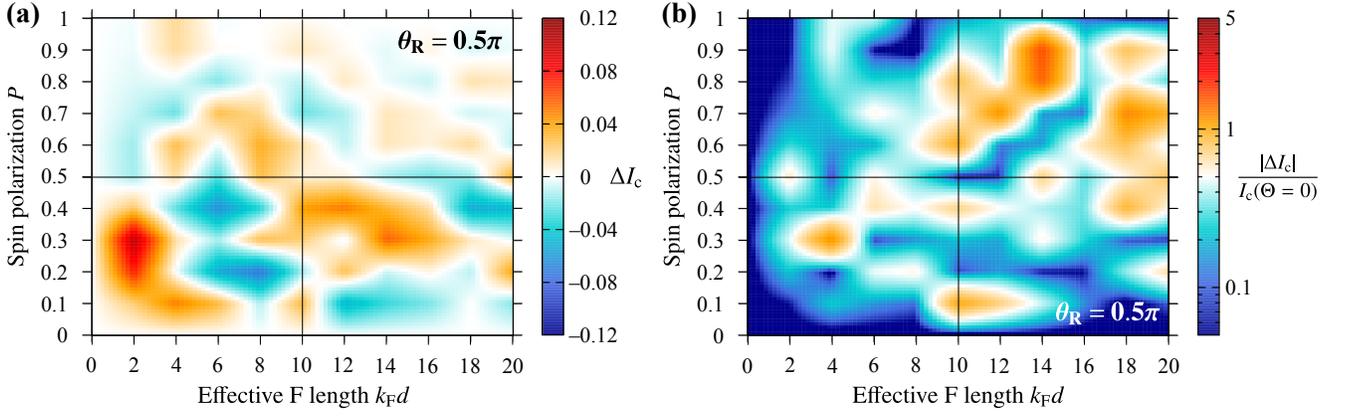

FIG. S2. (a) Critical-current difference $\Delta I_c = I_c^+ - |I_c^-|$—normalized as in the main text—as a function of the effective F length $k_F d$ and its spin polarization $P$ for magnetization along $+\hat{z}$ (i.e., $\Theta = 0.5\pi$) and Rashba angle $\theta_R = 0.5\pi$. (b) Corresponding relative SDE efficiency $|\Delta I_c|/I_c(\Theta = 0)$, where $I_c(\Theta = 0)$ is the polarity-independent (positive) critical current for in-plane magnetization along $+\hat{y}$.

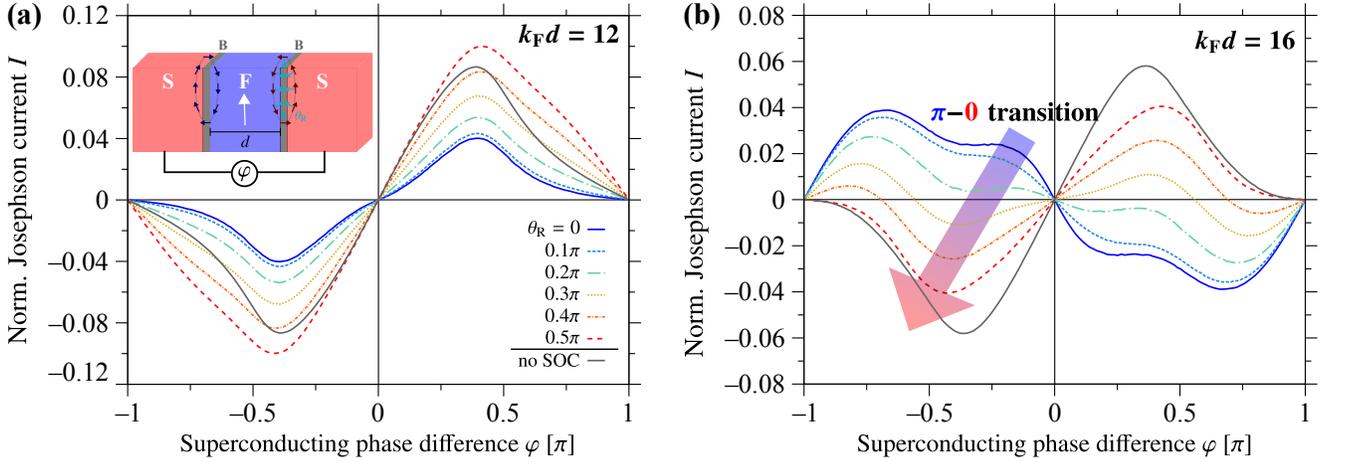

FIG. S3. (a) CPRs $I(\varphi)$—normalized as in the main text—of the vertical S/F/S Josephson junction for effective F lengths (a) $k_F d = 12$ and (b) $k_F d = 16$, and for various indicated Rashba angles $\theta_R$. The F magnetization is aligned along the $+\hat{y}$-in-plane direction (i.e., $\Theta = 0$ and $\Phi = 0.5\pi$) and the Rashba strength is $\lambda_R = 2m\alpha/\hbar^2 = 1$. For $k_F d = 16$, a more dominant RR component (larger $\theta_R$) results in a current-reversing $\pi$–0 transition.

as shown in the inset of Fig. S1(b). The corresponding, maximal-amplitude (due to the fully perpendicular magnetization), critical-current difference $\Delta I_c = I_c^+ - |I_c^-|$ as a measure of the USDE depends (nearly perfectly) sinusoidally on the Rashba angle $\theta_R$ with small deviations in the vicinity of $\theta_R = 0.05\pi$, which could be a reminiscent feature of the $\pi$–0-like back-transition.

In the main text, we ascribed the USDE to polarity- and field-orientation-dependent transmission probabilities that microscopically originate from precessions of the in-plane-polarized electron spins when traversing the F link. If the SOC at the left interface is of the CR and that at the right interface of the RR type, the transmission probabilities of (left-) right-going electrons depends on the angle $(\phi_{RR})$ $\bar{\phi}_{RR}$ that the precessing spins enclose with the preferred direction of the SOC when arriving at the second interface according to $\mathcal{T}_{RR}^{\leftarrow}(m_z > 0) \propto \cos^2(\phi_{RR}/2)$ and $\mathcal{T}_{RR}^{\rightarrow}(m_z > 0) \propto \cos^2(\bar{\phi}_{RR}/2)$—assuming $m_z > 0$ for the F magnetization. Depending on how fast the spins precess, either $\phi_{RR}$ or $\bar{\phi}_{RR}$ could lead to the larger transmission probability suggesting that the critical-current difference $\Delta I_c = I_c^+ - |I_c^-|$ can reverse its sign. The most convenient knobs to tune the precession are the length $d$ of the F or its spin polarization $P$. Figure S2(a) presents $\Delta I_c$ as a function of $d$ and $P$ for out-of-plane magnetization along $+\hat{z}$ ($\Theta = 0.5\pi$) and Rashba angle $\theta_R = 0.5\pi$ (pure RR SOC to induce the maximal USDE). We clearly observe the aforementioned tunability of $\Delta I_c$ (the USDE)—including the expected sign reversals—with $d$ and $P$, agreeing well with our spin-precession picture. The absolute maximum of $\Delta I_c$ occurs for a very thin F link of length $k_F d = 2$ and a weak spin polarization of $P = 0.3$. Figure S2(b) shows the relative SDE measure $|\Delta I_c|/I_c(\Theta = 0)$, normalizing $\Delta I_c$ to the polarity-independent in-plane-magnetization critical current $I_c(\Theta = 0)$, on a logarithmic scale. The maximal relative SDE efficiency reaches values of about 200 % at $k_F d = 14$ and $P = 0.9$, making the USDE a sizable effect.

For completeness, Fig. S3 presents the CPRs for the same parameters as in Fig. S1 but assuming that the magnetization is aligned along the in-plane $+\hat{y}$-direction ($\Theta = 0$ and $\Phi = 0.5\pi$). As mentioned above, the considered Rashba strength is not yet enough that CR SOCs switch the junction into the $\pi$-state at $k_\mathrm{F}d = 12$, panel (a). Nevertheless, the junction is close to a SOC-induced $0$–$\pi$ transition and the critical current is already strongly suppressed compared to the case without SOC. The critical current increases again within the $0$-state once the SOC at the right interface acquires a radial component. At the slightly larger $k_\mathrm{F}d = 16$, panel (b), CR SOCs induce indeed a $\pi$-state and, similarly to the out-of-plane magnetization discussed above, a $\pi$–$0$-like transition back into the $0$-state emerges when the Rashba angle grows. As expected from our tunneling picture elaborated in the main text, in-plane magnetization does not give rise to the spin precessions required to produce the USDE. As a result, all CPRs for in-plane magnetization are perfectly point-symmetric with respect to zero phase difference and the critical currents are independent of their polarity.

## III. ALTERNATIVE REALIZATION: LATERAL S/F/S JOSEPHSON JUNCTIONS

As an alternative platform for the USDE, we consider the tight-binding model for the two-dimensional lateral (planar) S/F/S Josephson junction consisting of two $s$-wave S leads and a F with an out-of-plane magnetization. We assume that CR SOC is present in both S leads and RR SOC only in the right S lead. The Rashba angle in the right S leads is $\theta_\mathrm{R}$ and the system is schematically depicted in Fig. S4(a). The Hamiltonian of the junction is given by

$$\hat{\mathcal{H}} = \hat{\mathcal{H}}_\mathrm{S,L} + \hat{\mathcal{H}}_\mathrm{F} + \hat{\mathcal{H}}_\mathrm{S,R}, \tag{S23}$$

$$\hat{\mathcal{H}}_\mathrm{S,L} = -t \sum_{\langle i,j \rangle, \sigma} \hat{c}^\dagger_{i,\sigma} \hat{c}_{j,\sigma} - \mu \sum_{j,\sigma} \hat{c}^\dagger_{j,\sigma} \hat{c}_{j,\sigma} - \sum_j \left[ \Delta \hat{c}^\dagger_{j,\uparrow} \hat{c}^\dagger_{j,\downarrow} + \mathrm{h.c.} \right]$$
$$+ \, \mathrm{i}\alpha \sum_{\mu = x,y} \sum_{\langle i,j \rangle_\mu} \sum_{\alpha,\beta} (\vec{n}^\mathrm{CR}_\mu \cdot \vec{\sigma})_{\alpha\beta} \hat{c}^\dagger_{i,\alpha} \hat{c}_{j,\beta}, \tag{S24}$$

$$\hat{\mathcal{H}}_\mathrm{S,R} = -t \sum_{\langle i,j \rangle, \sigma} \hat{c}^\dagger_{i,\sigma} \hat{c}_{j,\sigma} - \mu \sum_{j,\sigma} \hat{c}^\dagger_{j,\sigma} \hat{c}_{j,\sigma} - \sum_j \left[ \Delta \hat{c}^\dagger_{j,\uparrow} \hat{c}^\dagger_{j,\downarrow} + \mathrm{h.c.} \right]$$
$$+ \, \mathrm{i}\alpha \cos(\theta_\mathrm{R}) \sum_{\mu = x,y} \sum_{\langle i,j \rangle_\mu} \sum_{\alpha,\beta} (\vec{n}^\mathrm{CR}_\mu \cdot \vec{\sigma})_{\alpha\beta} \hat{c}^\dagger_{i,\alpha} \hat{c}_{j,\beta}$$
$$+ \, \mathrm{i}\alpha \sin(\theta_\mathrm{R}) \sum_{\mu = x,y} \sum_{\langle i,j \rangle_\mu} \sum_{\alpha,\beta} (\vec{n}^\mathrm{RR}_\mu \cdot \vec{\sigma})_{\alpha\beta} \hat{c}^\dagger_{i,\alpha} \hat{c}_{j,\beta}, \tag{S25}$$

and

$$\hat{\mathcal{H}}_\mathrm{F} = -t \sum_{\langle i,j \rangle, \sigma} \hat{c}^\dagger_{i,\sigma} \hat{c}_{j,\sigma} - \mu \sum_{j,\sigma} \hat{c}^\dagger_{j,\sigma} \hat{c}_{j,\sigma} - m_z \sum_{j,\sigma} \sigma \hat{c}^\dagger_{j,\sigma} \hat{c}_{j,\sigma}, \tag{S26}$$

respectively, where $\hat{\mathcal{H}}_\mathrm{F}$ is the Hamiltonian for the F and $\hat{\mathcal{H}}_\mathrm{S,L}$ ($\hat{\mathcal{H}}_\mathrm{S,R}$) for the left (right) S lead. Thereby, $t$, $\mu$, $\alpha$, $\Delta$, and $m_z$ are the hopping amplitude, the chemical potential, the Rashba strength, the $s$-wave pairing potential, and the magnitude of the exchange field along the $\hat{z}$-direction, while $\sigma = (-)1$ for spin (down) up. For the S leads, $\vec{n}^\mathrm{CR}_x = (0,1,0)$, $\vec{n}^\mathrm{CR}_y = (-1,0,0)$ are the Rashba vectors for CR and $\vec{n}^\mathrm{RR}_x = (1,0,0)$, $\vec{n}^\mathrm{RR}_y = (0,1,0)$ are the ones for RR SOC, accordingly.

To obtain the Josephson current, we calculate a bond current in the F for each phase difference between the S leads [S4]. The bond current at temperature $T$ is given by

$$I(\varphi) = \frac{e k_\mathrm{B} T}{\hbar} \sum_{n=0}^{N_\mathrm{max}} \sum_{\substack{i \in L \\ j \in R}} \mathrm{Im} \left[ \mathrm{Tr}(\hat{H}_{ij} \check{G}_{ji}(\mathrm{i}\omega_n)) - \mathrm{Tr}(\hat{H}_{ji} \check{G}_{ij}(\mathrm{i}\omega_n)) \right], \tag{S27}$$

where $\hat{H}_{ij}$ [$\check{G}_{ij}(\mathrm{i}\omega_n)$] are the hopping matrix (retarded Green's function) from site $j$ to $i$ calculated using KWANT [S5], and $\omega_n = (2n+1)\pi k_\mathrm{B} T$ with integer $n$ correspond to the fermionic Matsubara frequencies at temperature $T$ analogously to our analytical model above. The value $N_\mathrm{max}$ gives the maximum cut-off frequency and is determined dynamically to ensure convergence of the current. In Eq. (S27), the sites $i$ and $j$ are taken within the regions $L$ and $R$—see the tight-binding representation in Fig. S4(b)—and the summation over $i$ and $j$ is performed only when these two sites are nearest neighbor. We use parts of the code provided in the Supplemental Material of Ref. [S6] to calculate the Green's function $\check{G}_{ij}(\mathrm{i}\omega_n)$. The system parameters are taken as $\mu/t = -1$, $\alpha/t = 0.4$, $\Delta/t = 0.1$, and $k_\mathrm{B}T/t = 0.01$ to investigate the exchange-field dependence of the USDE (in terms of $\Delta I_\mathrm{c}$) and the anomalous phase shift $\varphi_0$; the lattice spacing is $a = 0.4$ nm indicating $t = \hbar^2/(2ma^2) \approx 238$ meV for the hopping constant ($m$ is the free-electron mass).

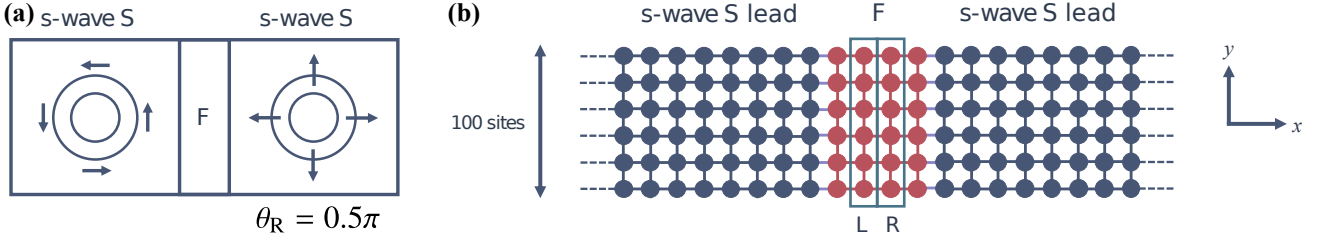

FIG. S4. (a) Schematical sketch of the considered planar S/F/S Josephson junctions consisting of two semi-infinite $s$-wave superconducting electrodes weakly coupled by a F link. In the left S, only CR SOC is present, while the right S hosts an admixture of CR and RR SOC quantified by the Rashba angle $\theta_R$ ($\theta_R = 0.5\pi$ is exemplarily shown). (b) Tight-binding representation of the system used for numerical implementation in KWANT assuming 100 sites along the transverse $\hat{y}$-direction and $N_F = 4$ sites along $\hat{x}$ inside the F; $L$ and $R$ indicate the cut-positions at which the Green's functions are evaluated (see text) to compute the Josephson CPRs.

As claimed in the main text, this lateral junction supports qualitatively similar physics as discussed for the vertical system. The CR SOC in the left S polarizes the electron spins along its preferred direction in the plane. When entering the F, the spins precess in-plane along the axis of the out-of-plane magnetization such that their transmission probabilities into the right S, which hosts also a nonzero RR component determined by the Rashba angle $\theta_R$, depend then on the relative angles between the precessing spin and the Rashba fields there. This mechanism results in transmission probabilities, and thereby critical currents, that are different for opposite directions and likewise also for opposite out-of-plane directions of the F magnetization—i.e., in the USDE. However, it is important to note that the current in the lateral junction flows in the plane of the SOC and the magnetization is perpendicular to this plane, whereas magnetization and current are parallel and perpendicular to the interfacial SOC in the vertical junction. This suggests that some transport features—particularly the $0-\pi$(-like) magnetizations effectively controllable through the interplay of SOC and magnetization in the vertical junction—can be different in the lateral system.

Figures S5(a) and S5(b) show the numerical KWANT results for the critical-current difference $\Delta I_c = I_c^+ - |I_c^-|$ for the Rashba angles $\theta_R = 0.25\pi$ and $\theta_R = 0.5\pi$, respectively. We clearly observe the USDE at finite out-of-plane magnetizations $m_z > 0$, displaying several sign reversals of $\Delta I_c$ and a generally increasing trend of the SDE with increasing $\theta_R$ similarly to the vertical junction. In contrast to the vertical case, the maximal values of $|\Delta I_c|$ occur in the half-metallic case when $m_z \to \mu$ (recall that $|\Delta I_c|$ oscillates as a function of $k_F d$ and $P$ in the vertical junction, Fig. S2) and its most pronounced sign reversal always coincides with a $0-\pi$(-like) transition [see the rapid $\varphi_0$-jumps in Figs. S5(c) and S5(d)]. Interestingly, this $0-\pi$(-like) transition is independent of $\theta_R$ and cannot be tuned by introducing more RR SOC in the right S. The reason for that is most likely the aforementioned elusive interplay between interfacial SOCs and magnetization in the vertical junction that is not present in the lateral case. As an important confirmation that the origin of the USDE is well-distinct from the finite Cooper-pair momentum, which has been identified as the key source of the conventional (in-plane-field) SDE [S7–S12], we also computed the spin-resolved Fermi surfaces inside the F link. The result [see the inset in Fig. S5(c)] indicates that the Cooper pairs do indeed not acquire a finite center-of-mass momentum (there is no shift or relative displacement of the Fermi surfaces w.r.t. each other) and another mechanism—i.e., the spin-precession tunneling—must generate the USDE.



\end{document}